\documentclass[10pt,a4paper]{article}
\usepackage[latin1]{inputenc}
\usepackage[T1]{fontenc}
\usepackage{amsmath,bm}
\usepackage{amsfonts}
\usepackage{amssymb}
\usepackage{subcaption}
\usepackage{graphicx}
\usepackage{mathtools}

\title{Pair Space in Classical Mechanics II. N-Body Central Configurations}
\author{Alon Drory \\ Afeka College of Engineering \\Tel-Aviv \\ alond@afeka.ac.il}

\begin{document}
	
	\maketitle
	
	\begin{abstract}
		
	\end{abstract}
	
	A previous work introduced pair space, which is spanned by the center of mass of a system and the relative positions (pair positions) of its constituent bodies. Here, I show that in the $N$-body Newtonian problem, a configuration that does not remain on a fixed line in space is a central configuration if and only if it conserves all pair angular momenta. For collinear systems, I obtain a set of equations for the ratios of the relative distances of the bodies, from which I derive some bounds on the minimal length of the line. For the non-collinear case I derive some geometrical relations, independent of the masses of the bodies. These are necessary conditions for a non-collinear configuration to be central. They generalize, to arbitrary $N$, a consequence of the Dziobek relation, which holds for $N=4$.
	
	\section{Introduction}
	\label{sec:intro}
	
	A previous work (hereafter referred to as PS-I) introduced a representation of classical mechanics in pair-space \cite{ps1}. Let a system contain $N$ bodies with masses $\{m_i\}$ and position vectors $\{ \bm{r}_i\}$ (in the usual geometric space). Pair space is spanned by the system's center of mass $ \bm{R}$ and the relative positions $\{ \bm{q}_{1 2},..., \bm{q}_{(N-1) N}\}$, which are defined as 
	\begin{equation}
		\label{pairq}
		\bm{q}_{ij} = \bm{r}_i - \bm{r}_j .
	\end{equation}

	Assume that the particles interact via pairwise potentials $v_{ij}( \bm{q}_{ij} )$. The system's potential energy is 
	\begin{equation}
		V=\sum\limits_{[i,j]} v_{ij}(\bm{q}_{ij})
	\end{equation}
	where $\sum_{[i,j]}$ means a sum over all pairs of distinct ordered indices with $i < j$. The system's kinetic energy is 
	\begin{equation}
		\label{kinetic}
		T = \frac{1}{2}M\bm{\dot{R}}^2 + \sum_{[i,j]}\frac{1}{2}\mu_{ij}\bm{\dot{q}_{ij}}^2 - \sum_{[i,j,k]}\frac{1}{2}\mu_{ijk}\left(\bm{\dot{q}_{ij}} + \bm{\dot{q}_{jk}} + \bm{\dot{q}_{ki}}\right)^2 ,
	\end{equation}
		where $\sum_{[i,j,k]}$ is a sum over all triplets of distinct ordered indices such that $i < j < k$. In this expression, $\mu_{ij}$ and $\mu_{ijk}$ are the pair and triplet reduced masses, respectively, defined as
	\begin{subequations}\label{reduced:main}
		\begin{align}
			\mu_{ij} &= \dfrac{m_i m_j}{M}    \label{reduced:a}\\  
			\mu_{ijk} &=\dfrac{m_i m_j m_k}{M^2} \label{reduced:b}
		\end{align}
	\end{subequations}

	Hamilton's principle does not apply to the Lagrangian $L = T-V$ because the pair positions are not all independent, verifying instead ``triangle conditions'', For every triplet of distinct indices $(i , j , k)$m we have that
	\begin{equation}
		\label{triangle}
		\bm{q}_{ij} + \bm{q}_{j k} + \bm{q}_{k i} = 0 .
	\end{equation}
	A configuration $\left\{\bm{q_{ij}}\right\}$ that verifies these conditions will be called \textit{\textbf{realizable}}. To ensure this condition, we introduce vector Lagrange multipliers, $\bm{\phi}_{ijk}$, defined for every triplet of monotonically ordered distinct indices, $i < j < k$. For convenience, we also define formal symbols where the ordering is different by 
	\begin{equation}
		\label{sgn}
		\bm{\phi}_{\sigma(i) \sigma(j) \sigma(k)} = sgn(\sigma)\bm{\phi}_{i j k}
	\end{equation}
	where $\sigma$ is a permutation of the indices $(i, j, k)$
	
	Hamilton's principle and the Euler-Lagrange equations hold for the pair Lagrangian, $L_{\pi}$, defined as
	\begin{equation}
		\label{lagrangianp}
		L _{\pi} = T  - V + \sum_{[i,j,k]}\bm{\phi}_{ijk} \left( \bm{q}_{ij} + \bm{q}_{j k} + \bm{q}_{k i}\right) 
	\end{equation}

	After some manipulation of the Euler-Lagrange equations, we can write the equations of motion in the following form \cite{ps1}:
	
	\begin{equation}
		\label{eqmotionq}
		\mu_{ij} \bm{\ddot{q}}_{ij}  +\frac{\partial v_{ij}(\bm{q}_{ij})}{\partial \bm{q}_{ij}} - \bm{J}_{ij} = 0
	\end{equation}
	for any pair of indices $i < j$, where we defined
	\begin{eqnarray}
		\label{jij}
		\bm{J}_{ij} = \sum_{\substack{n=1 \\n\neq i,j}}^{N}  \bm{\phi}_{ijn} 
	\end{eqnarray}
	Note that $\bm{J}_{ji}=- \bm{J}_{ij}$. 
	
	The explicit formula for these terms (see \cite{ps1}) is
	\begin{subequations} \label{eqmotionJ}
		\begin{align}
			&\frac{1}{\mu_{ij}}\bm{J}_{ij} = \sum_{\substack{k = 1 \\ k \neq i, j}}^{N} \dfrac{m_k}{M} \bm{F}_{ijk} \label{eqmotionJ:a}\\
			\text{where}  \nonumber \\
			&\bm{F}_{ijk} = 
			\frac{1}{\mu_{ij}}\frac{\partial v_{ij}(\bm{q}_{ij})}{\partial \bm{q}_{ij}} + \frac{1}{\mu_{jk}}\frac{\partial v_{jk}(\bm{q}_{jk})}{\partial \bm{q}_{jk}} + \frac{1}{\mu_{ki}}\frac{\partial v_{ki}(\bm{q}_{ki})}{\partial \bm{q}_{ki}} \label{eqmotionJ:b} 
		\end{align}
	\end{subequations}
	
	In particular, for the Newtonian potential, $v_{ij}(q_{ij}) = \dfrac{G M \mu_{ij}}{q_{ij}}$, we have that 
	\begin{equation}
		\label{newtonF}
		\bm{F}_{ijk} = G M \left( \frac{\bm{q}_{ij}}{q_{ij}^3} + \frac{\bm{q}_{jk}}{q_{jk}^3} + \frac{\bm{q}_{ki}}{q_{ki}^3} \right) 
	\end{equation}

	\section{Three-Body Energy and Angular Momentum}
	\label{sec:3bodymomentum}
	
	PS-I applied the pair space representation to the three-body problem, where three masses move under the influence of Newtonian potentials. For $N=3$ there is only one multiplier, $\bm{\phi} = \bm{\phi}_{123} = \bm{J}_{12} = \bm{J}_{23} = \bm{J}_{31}$.
	
	The equations of motion imply the conservation of energy in the form of contributions from pairs of bodies. The pair-energy is 
		\begin{equation}
		\label{energy}
		e_{ij} = \dfrac{1}{2} \mu_{ij} \vert \dot{\bm{q}}_{ij} \vert^2 - \dfrac{ G M \mu_{ij}}{q_{ij}}
	\end{equation}
	 Then from the equations of motion we obtain that the total pair-energy
	 \begin{equation}
	 	E_{\pi} =\sum_{\left[ i,j\right]} e_{ij}
	 \end{equation}
	 is conserved. This can be shown to be equal to the standard system's total energy minus the energy associated with the center of mass, which is decoupled from the other terms and is therefore independently conserved on its own. 
	 
	 Similarly, we define the pair-angular momentum associated with $\bm{q_{ij}}$ as 
	\begin{equation}
		\bm{L}_{ij}= \bm{q}_{ij} \times \mu_{ij}\bm{\dot{q}}_{ij}
	\end{equation}
	From the equations of motions, we obtain that the sum of all pair angular momenta is conserved. Furthermore, using Eq.(\ref{pairq}), we can express the total pair angular momentum through the particles' positions to obtain 
	\begin{equation*}
		\sum_{[i,j]}\bm{L}_{ij} = \sum_{p=1}^{3} \bm{r}_p \times \left(  m_p\bm{\dot{r}_p}\right) ,
	\end{equation*}
	where we assume that $\bm{r}_p$ are barycentric coordinates, i.e., $\sum_{p=1}^{3}m_p \bm{r}_p = 0$. This is equal to the system's standard angular momentum.
	
	Usually, individual pair angular momenta are not conserved, but we can consider particular solutions where they are. In PS-I, I showed that for $N=3$ these correspond to exactly two cases, the Euler collinear solution and the Lagrange equilateral solution. Pair space offer simple proofs of the existence of these solutions. Furthermore, it represents the Euler collinear solution differently from its usual form \cite{danby}. Here is a summary of the main results, as they will be useful in the next sections.
	
	\section{Newtonian Collinear Configuration}
	\label{sec:newtoncollinear}
	
	Given an ordering of the bodies along a line, we number them sequentially,, i.e. the mass at one end of the line is called $1$, the next one is $2$, followed by $3$. Since all pair vectors are collinear, they are expressible through a single parameter $\alpha$ as
	
	\begin{subequations}\label{coll1}
		\begin{align}
			\bm{q}_{23} (t) &= \alpha \bm{q}_{12}(t) ,  \label{coll1:a}\\
			\bm{q}_{31}(t) &= - \left[ 1 + \alpha\right]\bm{q}_{12}(t) , \label{coll1:b}
			\end{align}
	\end{subequations}
	There are two kinds of collinear motions. In the first, the masses move along a straight line fixed in space, either falling straight towards each other or else flying away. In the second case, $\alpha$ is constant (see proof in PS-I), and it is the root of the function
	
	\begin{equation}
		\label{Ex}
		E(x) =  M\left( x - \dfrac{1}{x^2} \right) + \left(m_1 - m_3 x \right)\left[ 1 + \dfrac{1}{x^2} - \dfrac{1}{\left(1 + x \right)^2}\right] .
	\end{equation}
	
	This is a monotonic function, with 
	\begin{subequations}  		\label{Elimits}
		\begin{align}
		E(x \to 0)   &\to - \infty  \\
		E(x \to \infty)  &\to \infty 
		\end{align}
	\end{subequations}
	Therefore, the function must have a single root. Multiplying Eq.(\ref{Ex}) by $x^2\left(1+x\right)^2$, we can see that its root $\alpha$ obeys a fifth order algebraic equation, which is equivalent to the standard Euler equation for the collinear solution \cite{danby, moulton1}. 
	
	In PS-I, I obtained several bounds for this root, in terms of the roots, $\sigma_k, \tau_k$ of the quartic equations
	\begin{subequations}
		\begin{align}
			\left( \dfrac{m_i + m_j}{m_k} \right)\sigma_k^2\left(1 + \sigma_k\right)^2 -1 - 2\sigma_k &= 0   , \\
		\tau_k^4 + 2 \tau_k^3 - \left( \dfrac{m_i + m_j}{m_k} \right)\left(1 + \tau_k\right)^2 &= 0  .
		\end{align}
	\end{subequations}
	Here, $(i,j,k)$ is some permutation of the indices $(1,2,3)$. In principle, $\sigma_k$ and $\tau_k$ can be written as closed-form expressions in the masses, but these are messy. Also, one of these equation suffices because $\sigma_k = 1/\tau_k$ (see PS-I). 
	
	The following bounds hold (see PS-I):
	
	1.  If $m_1 > \dfrac{4}{3}\left( m_3 + m_2\right) $, then $\alpha \in \left[ \sqrt{\dfrac{m_3}{m_1}} ,  \tau_1 \right] $    .
	
	2.  If $\dfrac{4}{3}\left( m_3 + m_2\right) \geq m_1 \geq m_3$, then $\alpha \in \left[\sqrt{\dfrac{m_3}{m_1}} ,  1 \right]$    .
	
	3. If $\dfrac{4}{3}\left( m_1 + m_2\right) \geq m_3 \geq m_1$, then $\alpha \in \left[ 1 , \sqrt{\dfrac{m_3}{m_1}}\right]$    .
	
	4.  If $m_3 > \dfrac{4}{3}\left( m_1 + m_2\right) $, then $\alpha \in \left[ \sigma_3 , \sqrt{\dfrac{m_3}{m_1}}\right]$   .
	
	\section{The $N$-body Case: Central Configurations}
	\label{sec:central}
	
	Looking now at the $N$-body case, the Euler solution is naturally extended to collinear configurations with more bodies \cite{moulton2}. The Lagrange solution, however, could be generalized in various ways, such as requiring that all bodies be equidistant, or else that they form the vertices of a regular polygon, or still other options. Pair space immediately suggests a natural and useful generalization by considering solutions of the $N$-body problem that conserve all the system's pair-angular momenta.
	
	For the general $N$-body problem, all pair- angular momenta $\bm{L}_{ij}$ are conserved if and only if, for all indices $(ij)$,
	\begin{equation*}
		0 = \dfrac{d \bm{L}_{ij}}{dt} = \mu_{ij} \bm{q}_{ij} \times \ddot{\bm{q}}_{ij}
	\end{equation*}
	Thus, there must be coefficients $\lambda_{ij} = \lambda_{ji}$ such that 
	\begin{equation*}
		\bm{\ddot{q}}_{i j} = - \lambda_{ij}\bm{q}_{i j}
	\end{equation*}
	The minus sign is introduced for future convenience.
	
	The triangle condition requires that for any three distinct indices $(i, j, k)$, we have
	
	\begin{equation}
		\label{suff}
		0 = \bm{\ddot{q}}_{ij} + \bm{\ddot{q}}_{j k} + \bm{\ddot{q}}_{k i} = - \lambda_{ij} \bm{q}_{ij} - \lambda_{jk}\bm{q}_{j k} - \lambda_{ki}\bm{q}_{k i}
	\end{equation}
	From the triangle condition itself, we can replace $\bm{q}_{k i}$ by $-\bm{q}_{i j}-\bm{q}_{j k}$. This yields
	\begin{equation*}
		0 = \left( \lambda_{ki} - \lambda_{ij} \right)  \bm{q}_{ij} + \left(\lambda_{ki} -  \lambda_{jk} \right) \bm{q}_{j k}
	\end{equation*}
	Taking the vector product of this equation once with $\bm{q}_{ij}$ and once with $\bm{q}_{j k}$ yields that
	\begin{equation}
		\label{ccsuff}
		0 = \left( \lambda_{ki} - \lambda_{ij} \right) \left(  \bm{q}_{ij} \times \bm{q}_{j k} \right) = \left(\lambda_{ki} -  \lambda_{jk} \right) \left(  \bm{q}_{ij} \times \bm{q}_{j k} \right)
	\end{equation}
	
	We shall prove the following
	
	\textbf{\underline{Proposition 1}:} Either all $\lambda_{\alpha \beta}$ are equal, for any pair of indices $(\alpha, \beta)$, or all the bodies are collinear.
	
	\textbf{Proof:} Select an arbitrary triplet of indices $(i , j , k)$. If $\bm{q}_{ij}$ and $\bm{q}_{j k}$ are not collinear, Eq.(\ref{ccsuff}) implies that 
	\begin{equation}
		\label{lambdacc}
		\lambda_{ij} =  \lambda_{jk} = \lambda_{ki} 
	\end{equation}
	
	Next suppose that $\left(  \bm{q}_{ij} \times \bm{q}_{j k} \right)= 0$. If every other body is also collinear with $(i , j , k)$, the proposition holds. Assume therefore that there is a body, e.g., $m_l$, that is not collinear with our triplet. Then consider the triplets $(i , j ,l)$, $(j , k , l)$ and $(i , k , l)$. By assumption, none of these are collinear. Therefore by Eq.(\ref{lambdacc}), we must have that
	\begin{eqnarray*}
		\lambda_{ij} = \lambda_{jl} =  \lambda_{li} \\
		\lambda_{jk} = \lambda_{kl} =  \lambda_{lj} \\
		\lambda_{ik} = \lambda_{kl} =  \lambda_{li} \\
	\end{eqnarray*}
	Combining these yields Eq.(\ref{lambdacc}) again. Hence, either all the bodies are collinear, or else for any triplet $(i,j,k)$, all the $\lambda$ coefficients must be identical. 
	
	Next let us take any two pairs of indices, $\alpha, \beta$ and $\gamma, \delta$. Consider the triplets $(\alpha, \beta, \epsilon)$, $(\alpha, \gamma, \epsilon)$ and $(\gamma, \delta, \epsilon)$, where $\epsilon$ is an arbitrary index. Then from the preceding argument,
	\begin{equation*}
		\lambda_{\alpha \beta} = \lambda_{\epsilon \alpha} = \lambda_{\gamma \epsilon} = \lambda_{\gamma \delta}  
	\end{equation*}
	Therefore, either all the bodies are collinear, or else all the proportionality coefficients are identical, for any pair of bodies in the system. QED.
	
\textbf{	\underline{Definition}:} An $N$-body configuration is called \textit{\textbf{central}} if for every index $i$,
	\begin{equation*}
		\ddot{\bm{r}}_i = - \lambda \left( \bm{r}_i - \bm{R} \right)  ,
	\end{equation*}
	where $\lambda$ is identical for all indices $i$ \cite{moeckel1}. 
	
	In the language of pair-coordinates, this implies that
	\begin{equation}
		\label{centraldef}
		\bm{\ddot{q}}_{i j} = - \lambda \bm{q}_{i j}
	\end{equation}
	
	Central configurations turn out to be important in the study of the $N$-body problem \cite{moeckel1, saari1, saari2,hampton1}. They are closely related, for example, to homographic solutions, which are solutions whose configuration remains self-similar at all times. Both the Euler and Lagrange solutions for three bodies are homographic. Pizzeti proved that homographic solutions are central at all times \cite{pizzeti}. 
	
	A motion is homothetic if the bodies converge along fixed straight lines until they collide. Homothetic solutions are also homographic, and form central configurations at all times, therefore. However, other initial configurations may also lead to $N$-body collisions (non-homothetically). Nevertheless, as the bodies approach each other, their orbits will asymptotically approach a central configuration \cite{siegel}. Conversely, when a system expands and the bodies separate from each other, it also tends towards a central configuration \cite{saari1}. Central configurations seem to play a fundamental role in the solution of the $N$-body problem, therefore.
	
	Proposition 1  shows that if all the pair-angular momenta are conserved, the motion of the system is either collinear or forms a central configuration. We can now prove:
	
	\textbf{\underline{Proposition 2}:} A non-collinear configuration is central if and only if it conserves all pair angular momenta.
	
	\textbf{Proof}. Proposition 1 implies one direction: if the configuration is not collinear and conserves all pair angular momenta, then there is a single parameter $\lambda$, such that for all indices $(i, j)$, $\bm{\ddot{q}}_{i j} = - \lambda \bm{q}_{i j}$. By Eq.(\ref{centraldef}), this means the configuration is central.
	
	Conversely, if a configuration is central, it verifies Eq.(\ref{centraldef}). Then the pair-angular momenta all have vanishing derivatives, since
	\begin{equation}
		\dfrac{d \bm{L}_{i j}}{dt} = \bm{q}_{ij} \times \mu_{ij}\bm{\ddot{q}}_{ij} = 0.
	\end{equation}
	
	QED.
	
	From the equations of motion, Eq.(\ref{eqmotionq}), the pair angular momenta are constant if and only if $\bm{q}_{ij} \times \bm{J}_{ij} = 0$. From Eq.(\ref{eqmotionJ}), this means that
	\begin{equation}
		\label{cccond}
		0 = \sum_{\substack{k = 1 \\ k \neq i, j}}^{N} m_k \bm{q}_{ij} \times \bm{F}_{ijk}
	\end{equation}
	For the Newtonian potential, we can substitute Eq.(\ref{newtonF}) into Eq.(\ref{cccond}) and use the triangle condition to replace $\bm{q}_{ki} = - \bm{q}_{ij} - \bm{q}_{jk}$. We obtain a set of $N(N-1)/2$ equations, one for each pair of (ordered) indices, which are necessary and sufficient conditions for a non-collinear realizable configuration to be central (as long as there are no collisions):
	\begin{equation}
		\label{cceq}
		\sum_{\substack{k = 1 \\ k \neq i, j}}^{N}  m_k \left(  \bm{q}_{ij} \times \bm{q}_{jk} \right)  \left(\frac{1}{q_{ik}^3} - \frac{1}{q_{jk}^3} \right) = 0
	\end{equation}
	
	The qualifier ''realizable'' in the previous sentence (i.e., a configuration that verifies the triangle conditions) is necessary. For example, suppose that $q_{ij} = q_{kl}$ for any two pairs of indices, i.e, all the bodies are equidistant from each other. In this case, every single term in the sums in Eq.(\ref{cceq}) vanishes. For $N$ bodies, this represents a regular simplex with $N$ vertices and such an object can only exists in a space of dimension $N-1$. Thus, for the physical three dimensional space, the only realizable cases are $N=3$ (the Lagrange equilateral triangle solution) and $N=4$, which is a regular tetrahedron. But we cannot arrange five bodies or more into a regular simplex in our physical space, so these solutions of Eqs.(\ref{cceq}) do not correspond to any realizable states because they violate the triangle conditions.
	
	\section{$N$-body Collinear Configurations}
	
	Let us consider now the collinear configuration of an $N$-body system. In this case, all the pair-positions are proportional to each other. First, we ask whether these proportions are constant in time. The argument here parallels the one in the three-body case (see PS-I). 
	
	Considering some pair position vector, $\bm{q}_{ij}$, there must be an $\alpha(t)$ such that, e.g., $\bm{q}_{ij} = \alpha(t)\bm{q}_{12}$. Take the vector product of $\bm{q}_{12}$ with the equation of motion of $\bm{q}_{ij}$. Clearly, $\bm{q}_{12} \times \bm{q}_{ij} = 0$. Since all the $\{\bm{{q}_{ij}}\}$ are collinear, Eq.(\ref{newtonF}) implies that
	\begin{equation*}
		\bm{q}_{12} \times \bm{F}_{abc} = 0  ,
	\end{equation*}
	for any triplet of indices $(a,b,c)$. Therefore, from Eq.(\ref{eqmotionJ}), we also have that 
	\begin{equation}
		\label{productij}
		\bm{q}_{12} \times \bm{J}_{ab} = 0   ,
	\end{equation}
	for any pair of indices $(a,b)$. 
	
	In particular, $\bm{q}_{12} \times \bm{J}_{12} = 0$. Therefore, the vector product of the equation of motion of $\bm{q}_{12}$ by $\bm{q}_{12}$ itself yields that
	\begin{equation}
		\label{ratio12}
		\bm{q}_{12} \times \ddot{\bm{q}}_{12} = 0   ,
	\end{equation}
	because all the other terms vanish. 
	
	Since Eq.(\ref{productij}) implies that $\bm{q}_{12} \times \bm{J}_{ij} = 0$ as well, we also have that
	\begin{equation*}
		\bm{q}_{12} \times \ddot{\bm{q}}_{ij} = 0   .
	\end{equation*}
	Now
	\begin{equation*}
		\bm{q}_{12} \times \ddot{\bm{q}}_{ij} = \bm{q}_{12} \times \left[  \ddot{\alpha} \bm{q}_{12} + 2\dot{\alpha}\dot{\bm{q}}_{12} + \alpha \ddot{\bm{q}}_{12}  \right]  .
	\end{equation*}
	The first and third term vanish, hence
	\begin{equation*}
		\dot{\alpha}\left[ \bm{q}_{12} \times \dot{\bm{q}}_{12}\right] = 0   .
	\end{equation*}
	
	As in the $N=3$ case, this equation has two solutions. In the first case, $\bm{q}_{12} \times \dot{\bm{q}}_{12} = 0$ and the bodies all move along a fixed straight line, some (possibly all) collapsing onto each other, and others (possibly none) flying away. The second solution is $\dot{\alpha} = 0$, wherein the proportion between any two pair positions is constant in time. We can now prove
	
	\textbf{\underline{Proposition 3}} If a collinear configuration does not remain on a fixed line at all times, it is central.
	
	\textbf{Proof}: By the above argument, if the motion of the bodies does not take place on a  fixed line, there must be a set of constants, $\left\lbrace a^{m,n}_{i,j} \right\rbrace $, such that
	\begin{equation*}
		\bm{q}_{i j} = a^{m,n}_{i,j} \bm{q}_{m n}
	\end{equation*}
	The constants are not all independent, but this is in keeping with the basic view of pair space, which seeks to take advantage of redundancy in order to simplify calculations and proofs. In particular, the following relations hold
	\begin{subequations}   \label{aijmn}
		\begin{align}
			&a^{m,n}_{i,j} = - a^{m,n}_{j, i} = - a^{n,m}_{i,j}    ,   \label{aijmn:a}\\
			&a^{m,n}_{p,q}	a^{p,q}_{i,j} = a^{m,n}_{i,j} ,    \label{aijmn:b} \\
			&a^{m,n}_{i,j} + a^{m,n}_{j,k} + a^{m,n}_{k,i} = 0   ,   \label{aijmn:c} 
		\end{align}
	\end{subequations}
	Note that there is no summing over $\left( p,q \right)$ in Eq.(\ref{aijmn:b}).	Eqs.(\ref{aijmn:a}) and (\ref{aijmn:b}) follow immediately from the definition of $a^{m,n}_{i,j}$, and Eq.(\ref{aijmn:c}) from the triangle condition.
	
	From Eq.(\ref{ratio12}), there must be a factor $\lambda$ such that (the minus sign is conventional)
	\begin{equation*}
		\ddot{\bm{q}}_{12} = -\lambda \bm{q}_{12}  .
	\end{equation*}
	Since for any pair $(i,j)$, $\bm{q}_{ij} = a^{1,2}_{i,j} \bm{q}_{1 2}$, we find that 
	\begin{equation*}
		\ddot{\bm{q}}_{i j} = a^{1,2}_{i,j} \left[ -\lambda \bm{q}_{1 2} \right]  = -\lambda \bm{q}_{i j}   .
	\end{equation*}
	This is the condition for a configuration to be central. QED
	
	Finally we can combine propositions 2 and 3. Together they immediately imply the following
	
	\textbf{\underline{Proposition 4:}} A realizable configuration that conserves all pair angular momenta individually either remains on a single fixed line at all times, or else it is a central configuration.
	
	\subsection{Equations for relative distances}
	\label{subsec:col_rel}
	
	For any of the $N!/2$ ordering of the masses on a straight line (starting from either end, hence the factor $1/2$), let us call the first mass $1$ then number the others sequentially so that $q_{12} < q_{13} < ... < q_{1N}$. As a result, $a^{m,n}_{i,j} > 0$ if the pairs $(m , n)$ and $(i , j)$ are both ordered in increasing order, or both in decreasing order.
	
	From Eq.(\ref{newtonF}), we obtain that
	\begin{subequations}
		\label{defh}  
		\begin{align}
			\bm{F}_{i j k} &= h^{m,n}_{i,j,k} \dfrac{G M}{ q_{m n}^3} \bm{q}_{m n} \label{defh:a}\\
			\text{where} & \nonumber \\
			h^{m,n}_{i,j,k} &= \dfrac{a^{m,n}_{i,j}}{\left| a^{m,n}_{i,j} \right|^3} +  \dfrac{a^{m,n}_{j,k}}{\left| a^{m,n}_{j,k} \right|^3} +  \dfrac{a^{m,n}_{k,i}}{\left| a^{m,n}_{k,i} \right|^3}.  \label{defh:b}
		\end{align}
	\end{subequations}
	
	This cumbersome notation is needed to take into account the possibility that the coefficients $a^{m,n}_{p,q}$ are negative. We shall be able to simplify the expressions for specific uses later on.
	
	Note that any permutation $\sigma$ of the indices $(i, j,k)$ changes at most the sign of these terms and
	\begin{equation*}
		h^{m,n}_{\sigma (i), \sigma(j), \sigma(k)} = sgn(\sigma) h^{m,n}_{i,j,k}   .
	\end{equation*}
	
	The terms $\bm{J}_{ij}$ in Eq.(\ref{eqmotionJ:a}) become
	\begin{equation}
		\frac{1}{\mu_{ij}}\bm{J}_{ij} = \left[ \sum_{\substack{k = 1 \\ k \neq i, j}}^{N} m_k  h^{m,n}_{i,j,k} \right] \dfrac{G }{q_{m n}^3} \bm{q}_{m n}   ,
	\end{equation}
	which is valid for any pair of indices $(m , n)$.
	
	In particular, the equation of motion of $\bm{q}_{m n}$ becomes
	\begin{equation}
		\label{motionqmn}
		\ddot{\bm{q}}_{m n} + \left[ M - \sum_{\substack{k = 1 \\ k \neq m, n}}^{N} m_k  h^{m,n}_{m,n,k} \right] \dfrac{G }{q_{m n}^3} \bm{q}_{m n} = 0    ,
	\end{equation}
	whereas that of any other pair position $\bm{q}_{ij} = a^{m,n}_{i,j}\bm{q}_{m n}$ is
	\begin{equation}
		\label{motionqij}
		a^{m,n}_{i,j}\ddot{\bm{q}}_{m n} + \left[ \dfrac{M a^{m,n}_{i,j}}{\left|a^{m,n}_{i,j} \right|^3 } - \sum_{\substack{s = 1 \\ s \neq i, j}}^{N} m_s  h^{m,n}_{i,j,s} \right] \dfrac{G }{q_{m n}^3} \bm{q}_{m n} = 0  .
	\end{equation}
	Isolating the expression of $\ddot{\bm{q}}_{m n}$ from Eq.(\ref{motionqmn}) and substituting it into Eq.(\ref{motionqij}), we obtain that
	\begin{equation}
		\label{collinear}
		\dfrac{M a^{m,n}_{i,j}}{\left|a^{m,n}_{i,j} \right|^3 } - \sum_{\substack{s = 1 \\ s \neq i, j}}^{N} m_s  h^{m,n}_{i,j,s} = a^{m,n}_{i,j}\left[ M - \sum_{\substack{k = 1 \\ k \neq m, n}}^{N} m_k  h^{m,n}_{m,n,k} \right]  .
	\end{equation}
	
	This is a set of equations for the coefficients $\left\lbrace a^{m,n}_{i,j} \right\rbrace $, i.e., for the ratios of distances between bodies in a collinear configuration. In order for the configuration to solve the equations of motions, these ratios must have specific values, therefore. Moulton has shown that there is a unique collinear configuration for every ordering of the masses along a line \cite{moulton2}, i.e., that there is a unique solution for this system of equations for any labeling of the masses (recall that here we number the masses consecutively along the line, and switching the positions of two masses means switching the values of the masses but not their indices).
	
	This mathematical representation of the collinearity condition differs from the standard one and offers therefore new opportunities to investigate this case. For example, we can derive some bounds on the extension of the line of bodies. 
	
	\subsection{Bounds on line length}
	\label{subsec:col_length}
	
	Let us choose $m=1, n = 2$, so that all distances are compared to the separation $q_{12}$. 
	For any $i < j$, we have that $a^{1,2}_{i,j} > 0$ and  Eq.(\ref{collinear}) takes the form
	\begin{equation}
		\label{collinE}
		a^{1,2}_{i,j}\left[M - \sum_{k=3}^N m_k h^{1,2}_{1,2,k} \right] - \frac{M}{\left(a^{1,2}_{i,j}\right)^2} + \sum_{\substack{s = 1 \\ s \neq i, j}}^{N} m_s  h^{1,2}_{i,j,s} = 0  .
	\end{equation}
	
		The distance to the last body gives the length of the line. Thus, we choose $i=2, j=N$. For succinctness, denote
	\begin{equation*}
		\beta = a_{2,N}^{1,2} = \frac{q_{2 N}}{q_{1 2}}  .
	\end{equation*}
	The length of the line of masses is 
	\begin{equation}
		\label{length}
		L = \left( 1 + \beta \right) q_{1 2}  .
	\end{equation}
	We shall now obtain a bound on this value.
	
	In Eq.(\ref{collinE}), we can separate the terms $k=N$ and $s=1$. The remaining terms in the sums can then be combined unto a single sum ranging from $3$ to $N-1$, and we obtain
	\begin{equation}
		\label{collinbeta}
		M\left(\beta - \frac{1}{\beta^2}\right) - \beta m_{N} h^{1,2}_{1,2,N} + m_1h^{1,2}_{2,N,1} + \sum_{k=3}^{N-1} m_k \left(h^{1, 2}_{2, N, k} - \beta h^{1,2}_{1, 2, k} \right)    = 0   .
	\end{equation}
	From the triangle condition,
		\begin{equation}
		\label{1plusbeta}
		a^{1,2}_{N,1} = -a^{1,2}_{1,N} = - \left[a^{1,2}_{1,2}+a^{1,2}_{2,N} \right] = - \left(1 + \beta \right)   .
	\end{equation}
		Then
	\begin{equation}
		h^{1,2}_{1,2,N} = h^{1,2}_{2,N,1} = 1 + \frac{1}{\beta^2} - \frac{1}{\left(1 + \beta\right)^2}  .
	\end{equation}
	
	Let us define the following function of $\beta$:
	\begin{equation}
		\label{ENdef}
		E_N\left(\beta\right) = M \left(\beta - \frac{1}{\beta^2}\right) + \left(m_1 - \beta m_N\right)\left[1 + \frac{1}{\beta^2} - \frac{1}{\left(1 + \beta\right)^2}\right] .
	\end{equation}
	Eq.(\ref{collinbeta}) then becomes
	\begin{equation}
		\label{ENeq}
		E_N\left(\beta\right) + \sum_{k=3}^{N-1} m_k \left(h^{1,2}_{2,N,k} - \beta h^{1,2}_{1,2,k} \right) = 0  .
	\end{equation}
	
	Since the masses are numbered sequentially, for any $k \leq N-1$ we have that $q_{2k} < q_{2N}$ and therefore $a^{1,2}_{2,k} <\beta$. For these same values of $k$, Eq.(\ref{defh:b}) becomes
		\begin{equation}
		h^{1,2}_{1,2,k} = \frac{1}{\left(a^{1,2}_{1,2}\right)^2} + \frac{1}{\left(a^{1,2}_{2,k}\right)^2} - \frac{1}{\left(a^{1,2}_{k,1}\right)^2} = 1 + \frac{1}{\left(a^{1,2}_{2,k}\right)^2} - \frac{1}{\left(1 + a^{1,2}_{2,k}\right)^2}.
	\end{equation}
	
	The last transition follows from a relation similar to Eq.(\ref{1plusbeta}), applied to $a^{1,2}_{k,1}$. Now the function $1 + \dfrac{1}{x^2} - \dfrac{1}{\left(1+x\right)^2}$ is monotonically decreasing from infinity (as $x \to 0$), to the value $1$ (as $x \to \infty$). Therefore, since $\beta > a^{1,2}_{2,k}$, we see that for every $k \ge 3$,
	\begin{equation}
		\label{boundh}
		h^{1,2}_{1,2,k} > 1 + \frac{1}{\beta^2} - \frac{1}{\left(1 + \beta\right)^2}   .
	\end{equation}
	
	Also, 
	\begin{equation}
		h^{1,2}_{2,N,k} = \frac{1}{\left(a^{1,2}_{2,N}\right)^2} - \frac{1}{\left(a^{1,2}_{N,k}\right)^2} - \frac{1}{\left(a^{1,2}_{k,2}\right)^2} = - \left[\frac{1}{\left(a^{1,2}_{2,k}\right)^2} + \frac{1}{\left(a^{1,2}_{k,N}\right)^2} - \frac{1}{\left(a^{1,2}_{2,k} + a^{1,2}_{k,N}\right)^2}\right],
	\end{equation}
	where the last equality follows from 
	\begin{equation}
		\label{1plusbeta:b}
		a^{1,2}_{2,N} = -a^{1,2}_{N,2} = - \left[a^{1,2}_{2,k} + a^{1,2}_{k,N} \right]   . 
	\end{equation}
	
	Clearly, $h^{1,2}_{2,N,k} \leq 0$ for every $k \ge 3$. Together with Eq.(\ref{boundh}), we now have that
	\begin{equation}
		- E_N\left(\beta\right)=\sum_{k=3}^{N-1} m_k \left(h^{1,2}_{2,N,k} - \beta h^{1,2}_{1,2,k} \right) \leq - \sum_{k=3}^{N-1} \beta m_k \left[ 1 + \frac{1}{\beta^2} - \frac{1}{\left(1 + \beta\right)^2} \right]  .
	\end{equation}
	
	From Eq.(\ref{ENeq}), we obtain the inequality
	\begin{equation}
	\label{eNineq}
	E_N^*(\beta) \geq 0   ,
	\end{equation}
	where
	\begin{equation}
	E^*_N(\beta) =  M \left(\beta - \frac{1}{\beta^2}\right) + \left(m_1 - \beta \sum_{k=3}^N m_k\right)\left[1 + \frac{1}{\beta^2} - \frac{1}{\left(1 + \beta\right)^2}\right]   .
	\end{equation}

	$E^*_N(\beta)$ is identical to the three-body function $E(\beta$), ( see Eq.(\ref{Ex})), but with $m_3$ replaced by an effective mass
	\begin{equation}
		\label{m3star}
		m_3^* = \sum_{k=3}^N m_k = M - m_1 - m_2  .
	\end{equation}
	
	In other words, $E^*_N$ represents a collapse of the N-body collinear configuration onto a three-body collinear configuration, in which the masses $m_k$ for $k \ge 3$ have been ''pushed out'' and merged with $m_N$.
	
	As a function of a general variable $x$, $E^*_N(x)$ is monotonically increasing from minus infinity to infinity (see section \ref{sec:newtoncollinear}), and has therefore a single positive root, which we denote $\beta^*$. Eq.(\ref{eNineq}) means that $E^*_N(\beta) \geq  E^*_N(\beta^*) = 0 $. The monotonicity of $E_N^*$ now implies that 
	\begin{equation}
		\label{betaineq}
		\beta \geq \beta^*
	\end{equation}
		
		From Eq.(\ref{length}), we now obtain the desired bound
		\begin{equation}
				\frac{L}{q_{12}} = \left( 1 + \beta \right)  \geq \left( 1 + \beta^* \right)
		\end{equation}

	Section \ref{sec:newtoncollinear} contains bounds on the value of $q_{23}/q_{12}$ for a three-body system. Eq.(\ref{betaineq}) then implies the following bounds for the $N$-body collinear configuration:
	
	1.  If $ 2 m_1 + m_2 \geq M $, then $\dfrac{L}{q_{12}} \geq 1 + \sqrt{\dfrac{m^*_3}{m_1}}  $    .
	
	2.  If  $\dfrac{7}{3}\left( m_1 + m_2\right) \geq M \geq 2m_1 + m_2$, then $\dfrac{L}{q_{12}} \geq 2$    .
	
	3.  If $M  > \dfrac{7}{3}\left( m_1 + m_2\right) $, then $\dfrac{L}{q_{12}} \geq 1 +  \sigma^*$ , where $\sigma^*$ is the solution of the quartic equation
	\begin{equation}
			\left( \dfrac{m_1 + m_2}{m^*_3} \right)\left( \sigma^*\right)^2\left(1 + \sigma^*\right)^2 -1 - 2\sigma^* = 0   .
	\end{equation}
	
	\subsection{Other bounds}
	\label{subsec:col_bounds}
	
	In the previous section, we mapped an $N$-body line onto a three body one by collapsing all the masses $\left\{m_k\right\}_{k=3}^{N-1}$ onto the mass $m_N$. Let us now consider a different case, in which the line is contracted by collapsing $\left\{m_k\right\}_{k=4}^N$ onto the mass $m_3$. We show that this yields an upper bound on the distance of $m_3$ from $m_2$ compared to the distance of $m_1$ from $m_2$. 
	
	We choose $i=2, j=3$ and denote
		\begin{equation*}
		\alpha = a_{2,3}^{1,2} = \frac{q_{23}}{q_{12}}  .
	\end{equation*}
		
	In Eq.(\ref{collinE}), we separate the term $k=3$ and $s=1$. The remaining terms in the sums can then be combined unto a single sum ranging from $4$ to $N$, and we obtain
	\begin{equation}
		\label{collinalpha}
		M\left(\alpha - \frac{1}{\alpha^2}\right) - \alpha m_3 h^{1,2}_{1,2,3} +m_1h^{1,2}_{2,3,1} + \sum_{k=4}^N m_k \left(h^{1,2}_{2,3,k} - \alpha h^{1,2}_{1,2,k} \right)    = 0   .
	\end{equation}
		
	As in Eq.(\ref{1plusbeta}), 
	
	\begin{equation}
		\label{1plusalpha}
		a^{1,2}_{3,1} = -a^{1,2}_{1,3} = - \left[a^{1,2}_{1,2}+a^{1,2}_{2,3} \right] = - \left(1 + \alpha \right)  .
	\end{equation}
	Hence
		\begin{equation}
		h^{1,2}_{1,2,3} = h^{1,2}_{2,3,1} = 1 + \frac{1}{\alpha^2} - \frac{1}{\left(1 + \alpha\right)^2}  .
	\end{equation}

	Similarly to $E_N$ as defined in Eq.(\ref{ENdef}), we define 
		\begin{equation}
		\label{E3def}
		E_3\left(\alpha\right) = M \left(\alpha - \frac{1}{\alpha^2}\right) + \left(m_1 - \alpha m_3\right)\left[1 + \frac{1}{\alpha^2} - \frac{1}{\left(1 + \alpha\right)^2}\right] ,
	\end{equation}
		 and rewrite Eq.(\ref{collinalpha}) as
		\begin{equation}
		\label{E3eq}
		E_3\left(\alpha\right) + \sum_{k=4}^N m_k \left(h^{1,2}_{2,3,k} - \alpha h^{1,2}_{1,2,k} \right) = 0 .
	\end{equation}
	
	Since we number the masses sequentially from one end of the line to the other, for any $k \geq 4$, we have that $q_{2k} \geq q_{23}$ and therefore $a^{1,2}_{2,k} \geq \alpha$. By an argument paralleling that which leads to Eq.(\ref{boundh}), we obtain that 
		\begin{equation}
		\label{boundh2}
		h^{1,2}_{1,2,k} \leq 1 + \frac{1}{\alpha^2} - \frac{1}{\left(1 + \alpha\right)^2}   .
	\end{equation}
	
	Also, 
		\begin{equation}
		h^{1,2}_{2,3,k} = \frac{1}{\left(a^{1,2}_{2,3}\right)^2} + \frac{1}{\left(a^{1,2}_{3,k}\right)^2} - \frac{1}{\left(a^{1,2}_{k,2}\right)^2} = \frac{1}{\alpha^2} + \frac{1}{\left(a^{1,2}_{3,k}\right)^2} - \frac{1}{\left(\alpha + a^{1,2}_{3,k}\right)^2},
	\end{equation}
	where the last equality follows from 
	\begin{equation}
		\label{1plusalpha:b}
		a^{1,2}_{k,2} = -a^{1,2}_{2,k} = - \left[a^{1,2}_{2,3} + a^{1,2}_{3,k} \right] = - \left(\alpha + a^{1,2}_{3,k}\right)  .
	\end{equation}
	
	Clearly, $h^{1,2}_{2,3,k} > 0$ for every $k \ge 4$. Together with Eq.(\ref{boundh2}), we now have that
	\begin{equation}
		-E_3\left(\alpha\right) = \sum_{k=4}^N m_k \left(h^{1, 2}_{2, 3, k} - \alpha h^{1,2}_{1, 2, k} \right) \geq - \sum_{k=4}^N \alpha m_k \left[ 1 + \frac{1}{\alpha^2} - \frac{1}{\left(1 + \alpha\right)^2} \right]  .
	\end{equation}
	
	From Eq.(\ref{E3eq}), we obtain the inequality
		\begin{equation}
		\label{e3ineq}
		E_3^*(\alpha)  \leq  0
	\end{equation}
	where
	\begin{equation}
		E_3^*(\alpha) =  M \left(\alpha - \frac{1}{\alpha^2}\right) + \left(m_1 - \alpha \sum_{k=3}^N m_k\right)\left[1 + \frac{1}{\alpha^2} - \frac{1}{\left(1 + \alpha\right)^2}\right]   .
	\end{equation}
		
	Once again, $E_3^*(\alpha)$ is identical to the three-body function $E(\alpha$)with $m_3$ being replaced by an effective mass
	\begin{equation}
		\label{m3starr}
		m_3^* = \sum_{k=3}^N m_k = M - m_1 - m_2  .
	\end{equation}
	
	Because this is a function of $\alpha$ rather than $\beta$, however, $E_3^*$ represents a collapse of the N-body collinear configuration onto a three-body collinear configuration in which the masses $m_k$ for $k \ge 4$ have been merged with $m_3$ instead of $m_N$ as before.
	
	The argument proceeds as before. Being monotonic, $E_3^*(x)$  has a single root, which we denote $\alpha^*$. Eq.(\ref{e3ineq}) means that $E_3^*(\alpha) \leq E_3^*(\alpha^*) = 0 $ and from the monotonicity of $E_3^*$, we obtain that 
	\begin{equation}
		\label{alphaineq}
		 \alpha \leq \alpha^* 
	\end{equation}

	This implies the following bounds for the $N$-body collinear configuration:
	
	1.  If $ \dfrac{7}{4}m_1 > M $, then $\alpha \leq \tau^* $    ,
	where $\tau^*$ is the solution of the quartic equation
	\begin{equation}
		\left(\tau^*\right)^4 + 2 \left(\tau^* \right)^3 - \left( \dfrac{m_2 + m^*_3}{m_1} \right)\left(1 + \tau^* \right)^2 = 0   .
	\end{equation}
	
	2.  If $2 m_1 + m_2 \geq M \geq \dfrac{7}{4}m_1$, then $\alpha \leq 1$    .
	
	3. If $ M \geq 2m_1 + m_2$, then $\alpha \leq \sqrt{\dfrac{m^*_3}{m_1}}$    .

	\section{Mass-Independent Relations}
	
	As described in section \ref{sec:central}, the necessary and sufficient conditions for a non-collinear configuration to be central are a set of vector equations, Eqs.(\ref{cceq}), one for each pair of indices $(i,j)$. Let us introduce the notations
	\begin{subequations} \label{AQ}
		\begin{align}
			A_{ikj} &= \dfrac{1}{q^3_{ik}} - \dfrac{1}{q^3_{kj}} = -A_{jki}    ,  \label{AQ:a} \\
			\bm{Q}_{ijk} &= \bm{q}_{ij} \times \bm{q}_{jk}  .\label{AQ:b}
		\end{align}
	\end{subequations}
	Eqs.(\ref{cceq}) become, for any pair of indices $(i,j)$,
	\begin{equation}
		\label{shortcceq}
		\sum_{\substack{k = 1 \\ k \neq i, j}}^{N}  m_k A_{ikj} \bm{Q}_{ijk}  = 0   .
	\end{equation}
	Using the triangle relation to replace $\bm{q}_{jk} = -\bm{q}_{ij} - \bm{q}_{ki}$ and $\bm{q}_{ij} = -\bm{q}_{jk} - \bm{q}_{ki}$, we see that $\bm{Q}_{ijk}$ is anti-symmetrical in its first two and last two indices: 

	\begin{subequations} \label{antisymQ}
		\begin{align}
		\bm{Q}_{ijk} &= -\bm{q}_{ji} \times \bm{q}_{jk} = \bm{q}_{ji} \times \bm{q}_{ki} = - \bm{Q}_{jik},  \label{symQ:a}  \\
		\bm{Q}_{ijk} &= -\bm{q}_{ki} \times \bm{q}_{jk} = -\bm{Q}_{ikj}.  \label{symQ:b}
		\end{align}
	\end{subequations}
	
	Applying the transpositions twice shows that $\bm{Q}_{ijk}$ is invariant under a cyclic permutation of its indices, i.e.,
	\begin{equation} \label{symQ}
	\bm{Q}_{ijk} = \bm{Q}_{jki} = \bm{Q}_{kij}  .
	\end{equation} 
	
	Consequently, we can rewrite Eq.(\ref{shortcceq}) more symmetrically as
	\begin{equation}
		\label{cceqfin}
		\sum_{\substack{k = 1 \\ k \neq i, j}}^{N}  m_k A_{ikj} \bm{Q}_{ikj}  = 0   .
	\end{equation}
These equations are not all independent, but that is of no concern for the present purpose.
	
	By taking the scalar product of Eqs.(\ref{cceqfin})  with an arbitrary vector $\bm{h}$, we obtain a set of scalar equations,
	\begin{subequations}   \label{cceqscalar}
		\begin{align}
		&\sum_{\substack{k = 1 \\ k \neq i, j}}^{N}  m_k A_{ikj} \overline{Q}_{ikj}  = 0,  \label{cceqscalar:a}  \\
\text{where}  \nonumber \\
		&\overline{Q}_{ikj} = \bm{Q}_{ikj} \cdot \bm{h}   .	\label{cceqscalar:b}
		\end{align}
	\end{subequations}

For a fixed index $j$, Eq.(\ref{cceqscalar:a}) can be put in matrix form as
\begin{subequations}   \label{matrixcc}
	\begin{align}
	&\sum_{\substack{\beta = 1 \\ \beta \neq \alpha, j}}^{N}  \Gamma^{j}_{\alpha \beta} m_{\beta} = 0   ,  \label{matrixcc:a}\\
\text{with}  \nonumber  \\
	&\Gamma^{j}_{\alpha \beta} = A_{\alpha \beta j} \overline{Q}_{\alpha \beta j} \label{matrixcc:b} .
	\end{align}
\end{subequations}
Since in Eq.(\ref{matrixcc:a}), $\alpha \neq \beta$, we define the diagonal terms to vanish, i.e.,
\begin{equation} \label{defgamma}
		\Gamma^{j}_{\alpha \alpha} \coloneq 0  .
\end{equation}

Eq.(\ref{matrixcc}) is a system of linear equations in the unknown masses $\{m_1 , m_2, \dots , m_N\}$. It has a nontrivial solution if and only if
\begin{equation}
	\label{detgamma}
	det \left(\Gamma^{j}_{\alpha \beta}\right) = 0
\end{equation}
for every index $j$. These are $N$ geometrical relations, independent of the masses, and they represent necessary (though not sufficient) conditions for a non-collinear configuration to be central. In other words, any shape that violates these relations cannot be a central configuration.

Consider first $N=3$ and choose $j=3$. Then, $\Gamma^3_{\alpha \beta}$ is a $2 \times 2$ matrix,
\begin{equation}
	\Gamma^3_{\alpha \beta} = \begin{pmatrix}
		0 & A_{1 2 3} \overline{Q}_{1 2 3} \\
		A_{2 1 3}  \overline{Q}_{2 1 3} & 0   .
	\end{pmatrix}
\end{equation}

Since the configuration is assumed non-collinear, $\bm{Q}_{\alpha \beta 3} \neq 0$. Since the vector $\bm{h}$ is arbitrary, $\overline{Q}_{\alpha \beta 3}$ will not vanish in general either. Hence we must have 
\begin{equation}
	A_{1 2 3}A_{2 1 3} = 0  \quad  \Longrightarrow q_{12}=q_{23} \quad \text{or} \quad  q_{12}=q_{13}   .
\end{equation}

Similarly, choosing $j=1$ and then $j=2$ yields, respectively, that 
\begin{subequations}
	\begin{align}
	q_{23}=q_{13} \quad &\text{or} \quad  q_{23}=q_{12} \\
	q_{13}=q_{23} \quad &\text{or} \quad  q_{13}=q_{12}    .
	\end{align}
\end{subequations}

Tracking all the possible combinations shows that ultimately we always have
\begin{equation}
	q_{12}=q_{23}=q_{13}  ,
\end{equation}
which represents the Lagrange equilateral solution. We see again that this is the  only non-collinear central configuration for $N=3$.

Next, consider $N=4$. First choose $j=2$. Now, $\Gamma^2_{\alpha \beta}$ is a $3 \times 3$ matrix, and with Eq.(\ref{defgamma}) we have that
\begin{equation}
	\Gamma^2_{\alpha \beta} = \begin{pmatrix}
		0 & A_{1 3 2} \overline{Q}_{1 3 2} & A_{1 4 2} \overline{Q}_{1 4 2}\\
		A_{3 1 2} \overline{Q}_{3 1 2} & 0 & A_{3 4 2} \overline{Q}_{3 4 2}\\
		A_{4 1 2} \overline{Q}_{4 1 2} & A_{4 3 2} \overline{Q}_{4 3 2} &0 
	\end{pmatrix}
\end{equation} 

The determinant of this matrix contains the products $\overline{Q}_{1 3 2}\overline{Q}_{4 1 2}\overline{Q}_{3 4 2}$ and $\overline{Q}_{1 4 2}\overline{Q}_{3 1 2}\overline{Q}_{4 3 2}$. Using Eqs. (\ref{antisymQ}) and (\ref{symQ}), we can replace $\overline{Q}_{4 1 2} = -\overline{Q}_{1 4 2}$ and so on, so that
\begin{equation}
\overline{Q}_{1 3 2}\overline{Q}_{4 1 2}\overline{Q}_{3 4 2} = - \overline{Q}_{1 4 2}\overline{Q}_{3 1 2}\overline{Q}_{4 3 2}
\end{equation}
 
 The condition of vanishing determinant then yields
 \begin{equation}
 	0 = det \left(\Gamma^{3}_{\alpha \beta}\right) = \left[A_{1 3 2}A_{4 1 2}A_{3 4 2} - A_{1 4 2}A_{3 1 2}A_{4 3 2}\right]\overline{Q}_{1 3 2}\overline{Q}_{4 1 2}\overline{Q}_{3 4 2}
 \end{equation}

Again, since the configuration is non-collinear, none of the vector products vanishes, and since the vector $\bm{h}$ is arbitrary, the product of the $\overline{Q}_{k l m}$'s will not vanish in general either. We conclude that 
\begin{equation} \label{origin-dziobek}
 A_{1 3 2}A_{4 1 2}A_{3 4 2} - A_{1 4 2}A_{3 1 2}A_{4 3 2} = 0
\end{equation}

Repeating the procedure with the remaining three values of $j$, we end up with four mass-independent relations that every non-collinear four-body central configuration must verify:
\begin{subequations} \label{dziobek}
	\begin{align}
		A_{3 2 1}A_{4 3 1}A_{2 4 1} - A_{4 2 1}A_{2 3 1}A_{3 4 1} &= 0 \label{dziobek:a} \\
		A_{1 3 2}A_{4 1 2}A_{3 4 2} - A_{1 4 2}A_{3 1 2}A_{4 3 2} &= 0 \label{dziobek:b} \\
		A_{4 1 3}A_{1 2 3}A_{2 4 3} - A_{2 1 3}A_{4 2 3}A_{1 4 3} &= 0 \label{dziobek:c} \\
		A_{3 1 4}A_{1 2 4}A_{2 3 4} - A_{2 1 4}A_{3 2 4}A_{1 3 4} &= 0 \label{dziobek:d} 
	\end{align}
\end{subequations}

Eq.(\ref{dziobek:b}) is equivalent to the well known Dziobek relation \cite{dziobek,corbera}, derived already in 1900, but here we have obtained three additional new ones. The set of Eqs.(\ref{detgamma}) for every value of $j$ can thus be considered an extension and generalization of the Dziobek relations for arbitrary values of $N$.

Note that these relations are necessary conditions for a configuration to be central, but not sufficient, as they do not restrict the values of the masses. However, they do suggest that the equality of mutual distances is an important criterion. The implications of such equalities for the case $N=4$ will be examined in a forthcoming work \cite{drory}.

\section{Summary}
	
	In a previous work, I have introduced the idea of pair space as a possible setting for classical mechanics and applied it to the three body Newtonian problem. This showed that the Lagrange and Euler solutions are distinguished by additional conserved quantities, the pair angular momenta. In the present work, I have considered the generalization of these configurations to the $N$-body problem. 
	
	The fundamental result is that non-collinear configurations conserve all pair angular momenta if and only if they are central configurations, while collinear configurations that conserve all pair angular momenta are central if and only if they do not remain on a fixed line throughout the motion. In other words, configurations that do not remain on a fixed line in space are central if and only if they conserve all pair angular momenta.
	
	This offers a different mathematical characterization of central configurations. For collinear configurations we obtain a set of equations that relate the various distances between bodies to each other. In particular, this yields certain bounds on the total length of the line with respect to the distance between its first two bodies as well as bounds on the distance of the third body from the second.
	
	For non-collinear central configurations, we obtained a set of algebraic equations that characterizes their shape. A detailed account of how these equations help map central configurations in the $N=4$ case is forthcoming. Here, we have obtained general geometrical relations, necessary but not sufficient, for a configuration to be central. The interest of these relations lies in their being independent of the masses. They serve therefore as a purely geometric classification of the shapes that central configurations \textit{might} take. In particular, any shape that violates these relations cannot be a central configuration, irrespective of the masses of the bodies involved.
	
	I hope that these results interest other researchers enough to make pair space a useful tool in future investigations.

\end{document}